\documentclass[preprint,aps]{revtex4}

\usepackage{amsmath}
\usepackage{graphicx}% Include figure files
\usepackage{dcolumn}% Align table columns on decimal point
\usepackage{bm}% bold math
\usepackage{epsfig}
\usepackage{graphics}
\usepackage{rotating}
\begin{document}

%\preprint{}

\title{Curved space and particle physics effects on the formation of Bose-Einstein condensation around a Reissner - Nordtstr{\o}m black hole}
%\hspace{20pt} \\
%\maketitle
\author{Recai Erdem}
\email{recaierdem@iyte.edu.tr}
\author{Bet{\"{u}}l Demirkaya}
\email{betuldemirkaya@iyte.edu.tr}
\author{Kemal G{\"u}ltekin}
\email{kemalgultekin@iyte.edu.tr}
\affiliation{Department of Physics \\
{\.{I}}zmir Institute of Technology\\
G{\"{u}}lbah{\c{c}}e, Urla 35430, {\.{I}}zmir, Turkey}

\date{\today}

\begin{abstract}
We consider two scalar fields interacting through a $\chi^*\chi\phi^*\phi$ term in the presence of a Reissner - Nordstr{\o}m black hole. Initially, only $\chi$ particles are present. We find that the produced $\phi$ particles are localized in a region around the black hole and have a tendency towards condensation provided that $\phi$ particles are much heavier than the $\chi$ particles. We also find that such a configuration is phenomenologically viable only if the scalars and the black hole have dark electric charges.

\end{abstract}
%\pacs{}

%\keywords{}

\maketitle

\section{introduction}

One of the most popular models of dark matter and dark energy are scalar field models where dark matter and dark energy are identified by scalar fields. Because of homogeneity and isotropy of the universe, these fields are taken to depend only on time at cosmological scales. This situation may be understood if the corresponding scalar fields form Bose-Einstein condensates at cosmological states. Therefore, there are many studies that study Bose-Einstein condensation of such scalar fields and their collapse at different cosmological and astronomical  backgrounds. Along the same lines, we had studied a model where initially only a scalar field $\chi$ is present and then it is converted to another scalar field $\phi$ through a $\chi^*\chi\phi^*\phi$ interaction term in the background of a Robertson-Walker metric \cite{Erdem-Gultekin}. We had shown that the evolution of $\phi$ is towards condensation provided $\phi$ particles are heavier than $\chi$ particles. In this study, we consider a similar setting in the background of a Reissner-Nordstr{\o}m black hole, and investigate the effect of geometry and the field content on the tendency of the system towards formation of a condensate. To be more specific, we assume that initially there is a homogeneous distribution of a diluted $\chi$ field in the presence of a Reissner-Nordstr{\o}m black hole \cite{Chandrasekhar}, and it transforms to a $\phi$ field, by time, through a $\chi^*\chi\phi^*\phi$ interaction term. We consider sufficiently early times of the process (so that the energy density of $\chi$ fields reaches a considerable value through superradiance \cite{superradiance} while the energy densities of the scalar fields do not reach sufficiently  high values to change the geometry appreciably). First, we study the motion of the scalar particles in the radial direction at the level of test particles. To this end, we mainly study the problem in the corresponding 1+1 dimensional subspace of the 3+1 dimensional space because we are mainly interested in the radial behaviours of fields. Then, we find an an approximate solution of the scalar field equations in 3+1 dimensions in closed form. We find that there are scalar field radial wave profiles that are soliton-like as expected from the analysis at the level of test particles.

In the next section, first, we review some basic well-known facts about the Reissner-Nordstr{\o}m metric that are essential for the derivation of our results,and provide the basic equations to be used in the next  section. In Section III we introduce a wave-like particular solution for the wave profile of charged scalar fields around a Reissner-Nordstr{\o}m black hole. In Section IV we discuss the phenomenological viability of this solution. Finally, in Section IV we conclude, and some technical details are derived in appendices.

\section{Framework}

The Reissner - Nordstr{\o}m metric is
\begin{equation}
ds^2\,=\,-f\,dt^2\,+\,f^{-1}dr^2\,+\, r^2\left(d\theta^2\,+\,\sin^2{\theta}\,d\varphi^2\right), \label{eq:1}
\end{equation}
where
\begin{equation}
f=\left(1\,-\,\frac{2M}{r}\,+\,\frac{Q^2}{r^2}\right).
\label{eq:2}
\end{equation}
Here $M$, $Q$ are the mass and the charge of the black hole, respectively. It describes a static black hole of mass $M$ and charge $Q$. One may either take the charge $Q$ to be a local $U(1)$ charge other than electric charge or one may take it to be a residual electric charge $Q\,\ll\,M$ (that may be due to much longer mean free path of an electron compared to nucleon in a hot baryonic plasma in a star, so the gravitational capture of some of the electrons by nearby astronomical objects before its collapse to form a black hole).

\subsection{Motion in radial direction}

Consider the following 1+1 dimensional subspace of (\ref{eq:1})
\begin{equation}
ds^2\,=\,-f\,dt^2\,+\,f^{-1}dr^2. \label{eq:3}
\end{equation}
The Lagrangian for a test particle of mass $m$ in the space given by (\ref{eq:3}) is
\begin{equation}
L\,=\,m\frac{ds}{d\tau}\,=\,m\sqrt{-f\dot{t}^2\,+\,f^{-1}\dot{r}^2}, \label{eq:4}
\end{equation}
where $\dot{t}=\frac{dt}{d\tau}$, $\dot{r}=\frac{dr}{d\tau}$ with $d\tau=\sqrt{-ds^2}$. The Lagrange equation for the coordinate $t$ results in conservation of h i.e. the total energy (including the potential energy) per unit mass of a test particle, namely
\begin{equation}
h\,=\,f\,\dot{t}\,=\,\frac{1}{m}\frac{\partial\,L}{\partial\,\dot{t}}\,=\,C\,=\,\mbox{constant}, \label{eq:5}
\end{equation}
where we have used
\begin{equation}
\left(-f\dot{t}^2+f^{-1}\dot{r}^2\right)\,=\,-1 \label{eq:6}
\end{equation}
for massive particles. Eq.(\ref{eq:6}) in combination with (\ref{eq:5}) results in
\begin{equation}
\dot{r}^2\,=\,C^2\,-\,\left(1\,-\,\frac{2M}{r}\,+\,\frac{Q^2}{r^2}\right).
\label{eq:7}
\end{equation}
As $r\,\rightarrow\,\infty$, (\ref{eq:7}) becomes
\begin{equation}
\dot{r}_\infty^2\,=\,C^2\,-\,1.
\label{eq:8}
\end{equation}

It is well-known that charged particles of charge $q$ (of the same charge as the black hole) with low enough energies $\omega$ with $0\,<\,\omega\,<q\frac{Q}{r_+}$ (where $r_+=M+\sqrt{M^2-Q^2}$ is the radius of the event horizon) can be scattered by Reissner-Nordstr{\o}m black holes. Moreover the scalar fields obeying the condition $m\,<\,\omega\,<\,q\frac{Q}{r_+}$ experience superradiance after being scattered \cite{RN-superradiance}.  Therefore one may consider scalar fields $\chi$ with mass $m_\chi$ and total energy $m_\chi\,C_\chi$ that fall to the black hole from large distance that may be approximated by infinity. Further one may consider another field $\phi$ with $m_\phi\,>\,m_\chi$ and a quartic interaction term $\chi^*\chi\phi^*\phi$ that results in $\chi\chi\,\rightarrow\,\phi\phi$ processes. Then, by conservation of energy (in the center of mass frame) we have $m_\chi\,C_\chi\,=\,m_\phi\,C_\phi$ i.e.
$C_\phi\,=\,C_\chi\frac{m_\chi}{m_\phi}$.  On the other hand, Eq.(\ref{eq:8}) implies that when the particle can barely reach infinity i.e. when  $\dot{r}_\infty^2\,=\,0$, $C^2\,=\,1$, and in general for a particle that can reach infinity $C^2\,\geq\,1$, and for a particle that cannot reach infinity for $C^2\,<\,1$ (if the particle is reflected by the black hole). These two results together imply that $\phi$ particles that are scattered by the black hole can reach only a finite distance from the black hole (which is the greater root of $(1-C^2)r_0^2-2Mr_0+Q^2=0$ for $C\,<\,1$, $Q\,<\,M$, that may be found by equating $\dot{r}$ in (\ref{eq:7}) to zero, the other root being inside the event horizon ) if $m_\chi\,<\,m_\phi$. In other words there will be belt of $\phi$ particles with zero or almost zero momenta around the black hole. This provides a suitable condition for formation of Bose-Einstein condensation. (In fact this explains why we do not consider the simpler case of a Schwarzschild black hole instead of a Reissner-Nordstr{\o}m black hole. In the case of a Schwarzschild black hole there will be no scattering from the horizon, so  there will be no $\chi\chi\,\rightarrow\,\phi\phi$ processes that are essential for the formation a belt of zero momenta scalar particles around the black hole that promotes formation of condensation.).
The conclusions that are derived above at the level of test particles above will be studied at the level of field theory in the following paragraphs.

\subsection{The field equations for the scalars}

We consider the following action for $\chi$ and $\phi$ particles
\begin{eqnarray}
S
&=&\int\;d^4x\,\sqrt{-g}\,\{-g^{\mu\nu}\left[D_\mu\phi\,\left(D_\nu\phi\right)^*\,+\,
D_\mu\chi\,\left(D_\nu\chi\right)^*\right]\,-\,m_\phi^2\left|\phi\right|^2\,-\,
m_\chi^2\left|\chi\right|^2\,-\,\lambda\,\phi^*\phi\chi^*\chi\}, \nonumber \\
&&\label{eq:9a}
\end{eqnarray}
where $D_\mu\,=\,\partial_\mu\,+\,iq\,A_\mu$ with $q$ being the electric charge of the scalar field and $A_\mu\,=\,\left(\frac{Q}{r}, 0, 0, 0\right)$ denoting the electric field of the black hole. We let both $\chi$ and $\phi$ have the same charge q. In (\ref{eq:9a}) we have neglected the effect of electromagnetic interactions between the scalar particles since the coupling constant of electromagnetic interactions is small, and the density of the scalar particles are taken to be small.

If the coupling term in (\ref{eq:9a}) is negligible with respect to the others, then the field equation for $\phi$ is
\begin{equation}
D_\mu\,D^\mu\,\phi\,-\,m_\phi^2\,\phi\,=\,0.
\label{eq:13}
\end{equation}
The corresponding equation for $\chi$ may be obtained by replacing $\phi$ in (\ref{eq:13}) by $\chi$. Using the ansatz \cite{RN-superradiance}
\begin{equation}
\phi_\omega\left(t,r,\theta,\varphi\right)\,=\,\sum_{l,m}e^{-i\omega\,t}Y_l^m\left(\theta,\varphi\right)\frac{\psi_\omega\left(r\right)}{r}
\label{c1}
\end{equation}
(\ref{eq:13}) reduces to
\begin{equation}
f^2\frac{d^2}{dr^2}\psi_\omega+ff^\prime\frac{d}{dr}\psi_\omega+\left[\left(\omega-\frac{qQ}{r}\right)^2-V\right]\psi_\omega\,=\,0,
\label{c2}
\end{equation}
where $\prime$ denotes derivative with respect to r, and
\begin{equation}
V= f\left(\frac{l(l+1)}{r^2}+\frac{f^\prime}{r}+m_\phi^2\right).
\label{c3}
\end{equation}

We seek an approximate solution of (\ref{c2}) for
\begin{equation}
\frac{f^\prime}{r}= \frac{2}{r^2}\left(\frac{M}{r}-\frac{Q^2}{r^2}\right)\,\ll\,m_\chi^2.
\label{c4}
\end{equation}
 In the next section we will show that (\ref{c4}) is satisfied for a wide range of $m_\chi^2$ provided that $r$ is not close to $r_+$. Note that (\ref{c4}) implies a similar relation for $m_\phi^2$ since $m_\chi\,<\,m_\phi$. For (\ref{c4}) (where  $m_\chi^2$ is replaced by $m_\phi^2$) and $l\,=\,0$ (i.e. for the motion that depends on r), (\ref{c2}) reduces to
\begin{equation}
\frac{d^2\psi_\omega}{dr_*^2}\,+\,\left[\left(\omega-\frac{qQ}{r}\right)^2\,-\,\tilde{m}_\phi^2\right]\,\psi_\omega\,=\,0,
\label{eq:15}
\end{equation}
where $dr_*\,=\,f^{-1}dr$, $\tilde{m}_\phi^2\,=\,f\,m_\phi^2$. In fact, (\ref{eq:15}) is similar to the corresponding exact 1+1 dimensional field equation (see Appendix A).
Eq.(\ref{eq:15}) may be also expressed as
\begin{equation}
\frac{d^2\psi_\omega}{dr_*^2}\,+\,\left[\omega^2\,-\,m_\phi^2\,-\,\frac{2(qQ\omega-m_\phi^2M)}{r}\,+\,\frac{q^2Q^2-m_\phi^2Q^2}{r^2}\right]\,\psi_\omega\,=\,0.
\label{eq:16}
\end{equation}

\section{a special solution}

\subsection{Derivation}

In the hope of obtaining a wave-like solution to (\ref{eq:16}) we consider a following type of solution
\begin{align}
\psi_\omega=e^{isr_*}g(r_*), \label{H1}
\end{align}
where $s^2=\omega^2-m^2$.
Eq.(\ref{eq:15}), after using (\ref{H1}), becomes
\begin{align}
\frac{g^{\prime\prime}}{g}+2is\frac{g^{\prime}}{g}-\frac{2(qQ\omega-m^2M)}{r}+\frac{(q^2-m^2)Q^2}{r^2}=0,\label{H2}
\end{align}
where $\prime$ denotes the derivative with respect to $r_*$.

We try the following choice
\begin{align}
\frac{g^{\prime}}{g}=\frac{\alpha_1}{r}+\frac{\beta_1}{r^2}+\frac{\gamma_1}{r^3}, \label{H3a}
\end{align}
\begin{align}
\frac{g^{\prime\prime}}{g}=\frac{\alpha_2}{r}+\frac{\beta_2}{r^2}+\frac{\gamma_2}{r^3}, \label{H3b}
\end{align}
where $\alpha_1$, $\beta_1$, $\gamma_1$, $\alpha_2$, $\beta_2$, $\gamma_2$ are some functions whose explicit forms will be derived below.
Hence, if such a solution exists, then (\ref{H1}) becomes
\begin{align}
\psi_\omega=e^{isr_*}\,\exp{\left[\int{\frac{\alpha_1\,r}{r^2-2Mr+Q^2}dr}+\int{\frac{\beta_1}{r^2-2Mr+Q^2}dr}+\int{\frac{\gamma_1}{r^3-2Mr^2+Q^2r}dr}\right]}.
\label{H1a}
\end{align}
We note this equation for reference later in the next subsection.

(\ref{H3a}) and (\ref{H3b}) solve (\ref{H2}) if (see Appendix B)
\begin{equation}
\alpha_1=\frac{9}{5}-\frac{9}{20}\bigg(\frac{M}{Q}\bigg)^2 \,, \ \ \ \ \ \ \beta_1=-\frac{9}{2}M \,, \ \ \ \ \ \ \gamma_1=3Q^2\,, \label{H5}
\end{equation}
\begin{align}
\alpha_2=0 \,, \ \ \ \ \ \ \beta_2=\bigg[\frac{9}{5}-\frac{9}{20}\bigg(\frac{M}{Q}\bigg)^2\bigg]\bigg[\frac{4}{5}-\frac{9}{20}\bigg(\frac{M}{Q}\bigg)^2\bigg] \,, \ \ \ \ \ \ \gamma_2=2\bigg[\frac{-18}{5}+\frac{63}{40}\bigg(\frac{M}{Q}\bigg)^2 \bigg]M.  \label{H6}
\end{align}
Inserting (\ref{H5}) and (\ref{H6}) into (\ref{H2}) and using $s^2=\omega^2-m^2$, we get three equations for three unknown quantities $\omega$, $q$, $m$ for a given $M$ and $Q$:
\begin{align}
(m^2-\omega^2)-\frac{M^2}{9Q^4}\bigg(\frac{18}{5}-\frac{63}{40}\frac{M^2}{Q^2}  \bigg)^2=0, \label{H7}
\end{align}
\begin{align}
-\frac{3M^2}{Q^2}\bigg(\frac{18}{5}-\frac{63}{40}\frac{M^2}{Q^2}\bigg)+\bigg(\frac{9}{5}-\frac{9}{20}\frac{M^2}{Q^2} \bigg) \bigg( \frac{4}{5}-\frac{9}{20}\frac{M^2}{Q^2} \bigg) +(q^2-m^2)Q^2=0, \label{H8}
\end{align}
\begin{align}
\frac{2M}{3Q^2}\bigg(\frac{18}{5}-\frac{63}{40}\frac{M^2}{Q^2}\bigg)\bigg(\frac{9}{5}-\frac{9}{20}\frac{M^2}{Q^2}\bigg)-2(qQ\omega-m^2M)=0. \label{H9}
\end{align}

Eq.(\ref{H7}) implies that $\omega\,\leq\,m$ (i.e. $s$ is imaginary or zero). In other words, the particles $\chi$ and $\phi$ are either, gravitationally bound or they have barely sufficient energy to come from infinity, which is in agreement with our assumptions about the $\chi$ and $\phi$ particles. The apparent independence of (\ref{H7}) of the charges and the masses of $\chi$ and $\phi$ may seem to imply independence of the form of the wave profile of the charges and the masses of $\chi$ and $\phi$ which may be misleading because the $M$ and $Q$ values in (\ref{H7}) are indirectly related to $m$ and $q$ through (\ref{H8}) and (\ref{H9}). Eq.(\ref{H1a}) after using (\ref{H7}) and (\ref{H5}) becomes
\begin{eqnarray}
\psi_\omega&=&
\exp\{ \left(\frac{\pm\,3 M\left(7 M^2-16 Q^2\right)}{40 Q^4 \sqrt{Q^2-M^2}}\right)
(\left(2 M^2-Q^2\right) \tan ^{-1}\left(\frac{r-M}{\sqrt{Q^2-M^2}}\right)\nonumber \\
&&+\left(Q^2-2 M^2\right) \tan ^{-1}\left(\frac{\text{r}_0-M}{\sqrt{Q^2-M^2}}\right) \nonumber \\
&&+\sqrt{Q^2-M^2} \left(M \log \left(-2 M r+Q^2+r^2\right)-M \log \left(-2 M \text{r}_0+Q^2+\text{r}_0^2\right)+r-\text{r}_0\right)\} \nonumber \\
 &&[\exp \{\left(-\frac{9 \left(M^2-4 Q^2\right)}{40 Q^2 \sqrt{Q^2-M^2}}\right)
 (2 M \tan ^{-1}\left(\frac{r-M}{\sqrt{Q^2-M^2}}\right)-2 M \tan ^{-1}\left(\frac{\text{r}_0-M}{\sqrt{Q^2-M^2}}\right) \nonumber \\
 &&+\sqrt{Q^2-M^2} \left(\log \left(-2 M r+Q^2+r^2\right)-\log \left(-2 M \text{r}_0+Q^2+\text{r}_0^2\right)\right))\} \nonumber \\
 &&-\frac{9M}{2\sqrt{Q^2-M^2}}\left(\tan ^{-1}\left(\frac{r-M}{\sqrt{Q^2-M^2}}\right)-  \tan ^{-1}\left(\frac{\text{r}_0-M}{\sqrt{Q^2-M^2}}\right)\right) \nonumber \\
 &&+\frac{6M}{2\sqrt{Q^2-M^2}}\left( -\tan ^{-1}\left(\frac{M-r}{\sqrt{Q^2-M^2}}\right)+ \tan ^{-1}\left(\frac{M-\text{r}_0}{\sqrt{Q^2-M^2}}\right)\right) \nonumber \\
 &&+3\sqrt{Q^2-M^2} (-\log \left(-2 M r+Q^2+r^2\right)+\log \left(-2 M \text{r}_0+Q^2+\text{r}_0^2\right) \nonumber \\
 &&+2 \log (r)-2 \log (\text{r}_0))].
 \label{H1b}
\end{eqnarray}
Depending on the relative values of $M$ and $Q$, the exponential function in (\ref{H1b}) is either an increasing or decreasing real exponential in $r$.
We discard the case of increasing real exponential functions because that case would correspond to an unphysical situation. We will discuss the implications of (\ref{H1b}) in the next section.

The  equations (\ref{H7}), (\ref{H8}), (\ref{H9}) may be solved for $m$, $\omega$, and $q$.
The solutions (that are obtained by Mathematica) as given below.
%\parbox{\textheight}{%
\begin{eqnarray}
m_{1,2}^2&=&\frac{9}{\left[(3200\, Q^{16} \left( Q^2-M^2\right)\right]}\{ 133\, M^6 Q^{10} + 124\, M^4 Q^{12} - 1104\, M^2 Q^{14} + 256\, Q^{16} \nonumber \\
&&\pm \;Q^{10}\,[182329\, M^{12} - 1244936\, M^{10} Q^2 + 3281072\, M^8 Q^4 -
      4249216\, M^6 Q^6 \nonumber \\
      &&+ 2552064\, M^4 Q^8 - 237568\, M^2 Q^{10} +
      65536\, Q^{12}]^\frac{1}{2}\},
\label{eq:m2}
\end{eqnarray}
\begin{eqnarray}
\omega_{1,2}&=&\mp\,\frac{3}{40 \,\sqrt{2}\,Q^8\left(Q^2-M^2\right)^\frac{1}{2}}\{98 M^8 Q^8-413 M^6 Q^{10}+1084 M^4 Q^{12}-1616 M^2 Q^{14} \nonumber \\
&&+256\,Q^{16}+Q^{10} [182329 M^{12}-1244936 M^{10} Q^2+3281072 M^8 Q^4-4249216 M^6 Q^6\nonumber \\
&&+2552064 M^4 Q^8 -237568 M^2 Q^{10}+65536 Q^{12}]^\frac{1}{2}\}^\frac{1}{2},
\label{eq:w2a}
\end{eqnarray}
\begin{eqnarray}
w_{3,4}&=&\mp\,\frac{3}{40 \,\sqrt{2}\,Q^8\,\sqrt{\left(Q^2-M^2\right)}}\{-98 M^8 Q^8+413 M^6 Q^{10}-1084 M^4 Q^{12}+1616 M^2 Q^{14} \nonumber \\
&&-256\,Q^{16}+Q^{10} [182329 M^{12}-1244936 M^{10} Q^2+3281072 M^8 Q^4-4249216 M^6 Q^6 \nonumber \\
&&+2552064 M^4 Q^8-237568 M^2 Q^{10}+65536 Q^{12}]^\frac{1}{2}\}^\frac{1}{2},
\label{eq:w2b}
\end{eqnarray}
\begin{eqnarray}
q_{1,2}&=&\mp\,\left[
\frac{3}{80 \sqrt{2} Q^{17}\, \sqrt{\left(Q^2-M^2\right)}\left(49 M^7-182 M^5 Q^2-8 M^3 Q^4+384 M Q^6\right)}\right] \nonumber \\
&&\times\,\{[49 M^6 Q^{10}-1100 M^4 Q^{12}+2384 M^2 Q^{14}-256 Q^{16}+Q^{10}[182329 M^{12}
-1244936 M^{10} Q^2 \nonumber \\
&&+3281072 M^8 Q^4-4249216 M^6 Q^6
+
2552064 M^4 Q^8-237568 M^2 Q^{10}+65536 Q^{12}]^\frac{1}{2}] \nonumber \\
&&\times\,[98 M^8 Q^8-413 M^6 Q^{10}+1084 M^4 Q^{12}-1616 M^2 Q^{14}++256 Q^{16} \nonumber \\
&&+ Q^{10}[182329 M^{12}-1244936 M^{10} Q^2+3281072 M^8 Q^4-4249216 M^6 Q^6 \nonumber \\
&&+2552064 M^4 Q^8-237568 M^2 Q^{10}+65536 Q^{12}]^\frac{1}{2}]^\frac{1}{2} \},
 \label{eq:q12}
\end{eqnarray}
\begin{eqnarray}
q_{3,4}&=&\pm\,
\left[
\frac{3}{80 \sqrt{2}\, Q^{17}\, \left(49 M^7-182 M^5 Q^2-8 M^3 Q^4+384 M Q^6\right)\sqrt{\left(M^2-Q^2\right)}}\right] \nonumber \\
&&\times\,\{[-98 M^8 Q^8+413 M^6 Q^{10}-1084 M^4 Q^{12}+1616 M^2 Q^{14}-256 Q^{16} \nonumber \\
&&+Q^{10}\,[182329 M^{12}-1244936 M^{10} Q^2+3281072 M^8 Q^4-4249216 M^6 Q^6+2552064 M^4 Q^8 \nonumber \\
&&-237568 M^2 Q^{10}+65536 Q^{12}]^\frac{1}{2}]^\frac{1}{2}
[-49 M^6 Q^{10}+1100 M^4 Q^{12}-2384 M^2 Q^{14} +256 Q^{16}\nonumber \\
&&+Q^{10} [182329 M^{12}-1244936 M^{10} Q^2+3281072 M^8 Q^4-4249216 M^6 Q^6 \nonumber \\
&&+2552064 M^4 Q^8-237568 M^2 Q^{10}+65536 Q^{12}]^\frac{1}{2}]\}.
 \label{eq:q22}
\end{eqnarray}

\subsection{Viability of the Solution}

To check the viability of the solution, it is more suitable to express $r_*$ in (\ref{eq:15}) or (\ref{eq:16}) in terms of multiples of $M$ or the mass of the sun ($M_\circ$). Then, for example, (\ref{eq:15}) may be expressed as (see Appendix C)
\begin{align}
\frac{d^2\psi_\omega}{d\overline{r}_*^2}+\bigg[\big(\overline{\omega}-\frac{\overline{q}\overline{Q}}{\overline{r}}\big)^2-\overline{m}_\phi^2\left(1-\frac{2\overline{M}}{\overline{r}} + \frac{\overline{Q}^2}{\overline{r}^2}\right) \bigg]\psi_\omega=0.
\label{scaled_eq}
\end{align}
Here
\begin{align}
\overline{r}_*=\frac{c^2}{GM_\circ}r_*,~~ \overline{r}=\frac{c^2}{GM_\circ}r, \label{bar1}
\end{align}
\begin{align}
&\overline{\omega}=\omega\frac{GM_\circ}{c^3}=\bigg(\frac{\omega}{s^{-1}}\bigg)5\times 10^{-6}, \label{bar2}
\\&
\overline{m}=m\frac{GM_\circ}{\hbar c} =\bigg( \frac{m}{kg}\bigg)\,4.5\times 10^{45}, \label{bar3}
\\&
\overline{M}=\frac{M}{M_\circ}=\bigg(\frac{M}{kg}\bigg)2\times 10^{-30}, \label{bar4}
\\&
\overline{Q}=\frac{Q\sqrt{k}}{M_\circ \sqrt{G}}=\bigg(\frac{Q}{C}\bigg)5,7\times 10^{-21}, \label{bar5}
\end{align}
\begin{align}
\overline{q}=q\frac{M_\circ \sqrt{Gk}}{\hbar c}=\bigg(\frac{q}{C}\bigg) 4,8 \times 10^{55}, \label{bar6}
\end{align}
where $k$ is the Coulomb's constant and we have explicitly written $G$, $c$, $\hbar$, $k$ (that we had set to 1) to see the phenomenological contents of these quantities.

$\bar{m}_{\chi(\phi)}^2$ and $\bar{m}_{\chi(\phi)}^2$ should be positive real numbers. This condition restricts possible values of $\bar{Q}$ for a given value of $\bar{M}$.  To this end, first we determine the roots of $\bar{m}_{1,2}^2=0$ in the equation obtained from (\ref{eq:m2}) by replacing the quantities in (\ref{eq:m2}) by their barred forms. We find the real roots of (\ref{eq:m2}) as $\bar{Q}_{1,2}=\pm\,\frac{\sqrt{7}\bar{M}}{4}$. Then, we plot $\bar{m}_{1,2}^2$ versus $\bar{Q}$ graphs for various values of $\bar{M}$. We find that $\bar{m}_{1}^2$ is real and positive in the intervals $\bar{Q}\,<\,-\frac{\sqrt{7}\bar{M}}{4}$ and $\bar{Q}\,>\,\frac{\sqrt{7}\bar{M}}{4}$. On the other hand, $\bar{m}_{2}^2$ is real and positive in the intervals $-\frac{\bar{M}}{2} \sqrt{\frac{1}{2} \left(\sqrt{697}-23\right)}\,<\,\bar{Q}\,<\,\frac{\bar{M}}{2} \sqrt{\frac{1}{2} \left(\sqrt{697}-23\right)}$ and $\frac{\sqrt{7}\bar{M}}{4}\,<\,\bar{Q}\,<\,\bar{M}$, $-\bar{M}\,<\,\bar{Q}\,<\,-\frac{\sqrt{7}\bar{M}}{4}$.

We have also checked the consistency of the formulation by solving the equations (\ref{H7}), (\ref{H8}), (\ref{H9}) (where all quantities are replaced by their barred forms) for
$\bar{m}^2$, $\bar{\omega}$, $\bar{Q}$ for $\bar{q}=0$. We have found that the corresponding $\bar{Q}$ and $\bar{M}$ satisfy $\bar{Q}=\pm\frac{\bar{M}}{4} \sqrt{\frac{1}{10} \left(299+\sqrt{52681}\right)}$) (which corresponds to the case $\bar{m}=\bar{m}_{2}$). As expected from the discussion in the preceding paragraph we find that it gives $\bar{q}=0$ for $\bar{m}_{2}^2\,<\,0$. Therefore, we conclude that there are no physical solutions for $q=0$. In other words, the solution described in the paper is realized only for $q\,\neq\,0$.

 Next, we discuss the order of the values of $\bar{m}^2$ and $\bar{q}$ for the phenomenologically viable intervals of $\bar{Q}$ discussed above. There are four relevant sets of parameters, namely, $\{$$|\bar{Q}|\,>\,\frac{\sqrt{7}\bar{M}}{4}$, $\bar{m}_1^2$, $\bar{\omega}_1$, $\bar{q}_1$$\}$, $\{$$|\bar{Q}|\,>\,\frac{\sqrt{7}\bar{M}}{4}$, $\bar{m}_1^2$, $-\bar{\omega}_1$, $-\bar{q}_1$$\}$, $\{$$|\bar{Q}|\,<\,\frac{\bar{M}}{2} \sqrt{\frac{1}{2} \left(\sqrt{697}-23\right)}$ or $\frac{\sqrt{7}\bar{M}}{4}\,<\,|\bar{Q}|\,<\,\bar{M}$, $\bar{m}_2^2$, $\bar{\omega}_3$, $\bar{q}_3$$\}$, $\{$$|\bar{Q}|\,<\,\frac{\bar{M}}{2} \sqrt{\frac{1}{2} \left(\sqrt{697}-23\right)}$ or $\frac{\sqrt{7}\bar{M}}{4}\,<\,|\bar{Q}|\,<\,\bar{M}$, $\bar{m}_2^2$, $-\bar{\omega}_3$, $-\bar{q}_3$$\}$ where the subindices are the ones in (\ref{eq:m2})-(\ref{eq:q22}). We have used a Mathematica code to try  values of $\bar{M}$ (as multiples of the mass of the Sun), and different values of $\bar{Q}$ to find the corresponding values of $\bar{q}_i$, $\bar{m}_i$. We observe that  (the positive) $\bar{m}_1^2$ values go to their maximum values (that decrease with increasing $\bar{M}$ and which is about 1 for $\bar{M}=1$) as $\bar{Q}\,\rightarrow\,\pm\bar{M}$ and goes to zero as $\bar{Q}\,\rightarrow\,\pm\frac{\sqrt{7}\bar{M}}{4}$, and takes intermediate values in between. This implies that the value of $\bar{m}_1^2$ can not exceed 1 that corresponds to a mass of order of $10^{-45}\,kg$ i.e. of the order of $10^{-9}\,eV/c^2$.  On the other hand  (the positive) $\bar{m}_2^2$ values go to plus infinity as $\bar{Q}\,\rightarrow\,0$ or as $\bar{Q}\,\rightarrow\,^-\bar{M}$ or as $\bar{Q}\,\rightarrow\,-^+\bar{M}$ while it tends to zero as $\bar{Q}\,\rightarrow\,\pm\,\frac{\bar{M}}{2} \sqrt{\frac{1}{2} \left(\sqrt{697}-23\right)}$, and takes all intermediate values in between. Note that positive real values of $\bar{m}_2^2$ are only possible for $|\bar{Q}|\,<\,\bar{M}$ i.e. for black holes. Once the values $\bar{m}_i$ are determined, the values of $\bar{\omega}_i$ may be determined by (\ref{H7}). In agreement with the the range of $\bar{Q}$ for positive values $\bar{m}_1^2$, $\bar{q}_{1,2}$ have non-trivial values for $|\bar{Q}|\,>\,\frac{\sqrt{7}\bar{M}}{4}$, and they go to $\pm$ infinity as $\bar{Q}\,\rightarrow\,\pm\,\frac{\sqrt{7}\bar{M}}{4}$ and go to zero as $\bar{Q}\,\rightarrow\,\pm\,\infty$ and $\bar{Q}\,\rightarrow\,\pm\,\frac{\bar{M}}{4}\sqrt{\frac{1}{10} \left(299-\sqrt{52681}\right)}$, and takes intermediate values (smaller than of order of 1) in between. Hence, the relevant set of parameters are $\{$$|\bar{Q}|\,>\,\frac{\sqrt{7}\bar{M}}{4}$, $\bar{m}_1^2$, $\pm\bar{\omega}_1$, $\pm\bar{q}_1$$\}$ and $\{$$\bar{M}\,>\,|\bar{Q}|\,>\,\frac{\sqrt{7}\bar{M}}{4}$, $\bar{m}_3^2$, $\pm\bar{\omega}_3$, $\pm\bar{q}_3$$\}$. To summarize, this analysis gives two main results. The first result is that the allowed values of $\bar{m}^2$ are smaller than $10^{-9}\,eV/c^2$. The second result is that $Q$ and $q$ in this study can not be the usual electric charge in the light of the condition $|\bar{Q}|\,>\,\frac{\sqrt{7}\bar{M}}{4}$ derived above and in the light of absence of observation of astronomical compact object with a significant value of an electric charge. This scenario is possible if we take $q$ and $Q$ as electric charges of a dark U(1) force \cite{dark-photon}. Another result of the above analysis is that the solutions with $|    \bar{Q}|\,<\,\frac{\sqrt{7}\bar{M}}{4}$ are unphysical.

\begin{figure}[h]
\centerline{\includegraphics[scale=0.6]{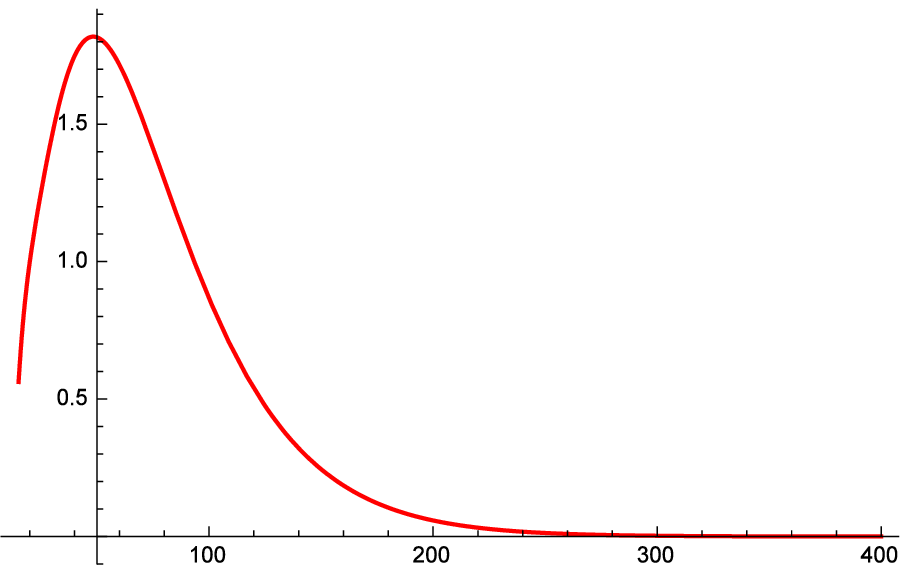}}
\caption{$\bar{r}$ versus $|\psi_\omega|$ graph for $\bar{M}=10$, $\bar{Q}=17$, $\bar{r}_0=20$ }
\label{fig:1}
\end{figure}

\begin{figure}[h]
\centerline{\includegraphics[scale=0.6]{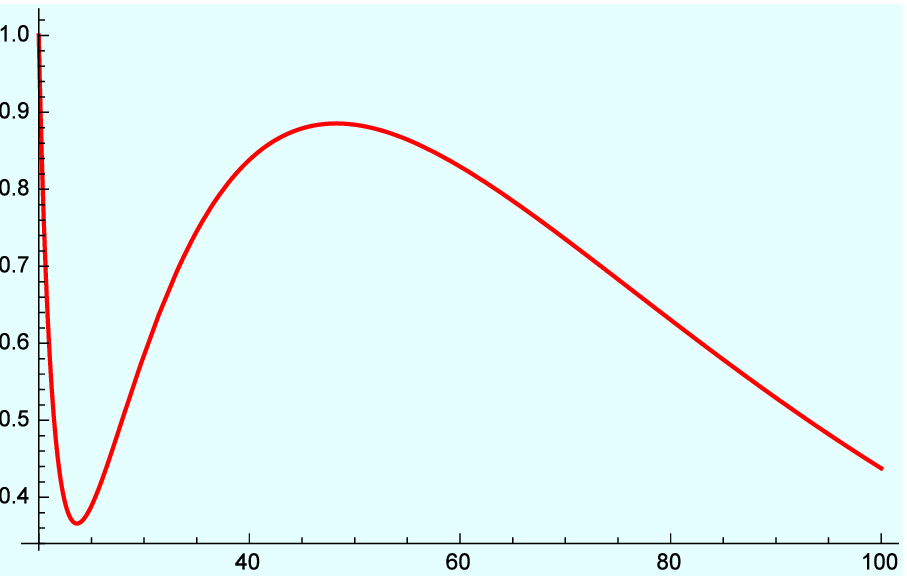}}
\caption{$\bar{r}$ versus $|\psi_\omega|$ graph for $\bar{M}=10$, $\bar{Q}=7$, $\bar{r}_0=20$ }
\label{fig:2}
\end{figure}

\begin{figure}[h]
\centerline{\includegraphics[scale=0.6]{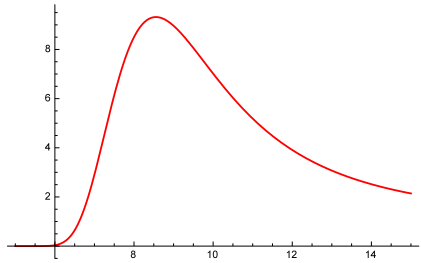}}
\caption{$\bar{Q}$ versus $|\psi_\omega|$ graph for $\bar{M}=10$, $\bar{r}=20$, $\bar{r}_0=30$. }
\label{fig:3}
\end{figure}

 Now we check if one may find the wave profiles that are expected from a test particle treatment i.e. if there exist wave profiles that peak about some values of $r$ as discussed in the first part of the preceding section. Given the complicated form of (\ref{H1b}) it is difficult to deduce simple general rules for the behaviour of $\psi_\omega$. We have plotted $\psi_\omega$ plots for various values of its parameters by using a Mathematica code. We have found mainly two types of behaviours for $|\psi_\omega|$, namely, exponentially increasing with increasing $r$, exponentially decreasing with increasing $r$ with a local peak. We discard the exponentially increasing ones since they are unphysical. Some examples of the physically relevant cases for the physically relevant set (that is discussed above) $|\bar{Q}|\,>\,\frac{\sqrt{7}\bar{M}}{4}$ are shown in the figures \ref{fig:1}, \ref{fig:2}. We find that the corresponding $\bar{M}=10$ and $\bar{Q}=7$, $\bar{Q}=17$ result in $-\bar{\omega}_1=0.0219237$, $-\bar{\omega}_1=0.0550123$, and it seems that all $\pm\bar{\omega}_1$ for $|\bar{Q}|\,>\,\frac{\sqrt{7}\bar{M}}{4}$ are real while all $\pm\bar{\omega}_1$ for $|\bar{Q}|\,<\,\frac{\sqrt{7}\bar{M}}{4}$ are imaginary (so, the corresponding solutions are unstable). Moreover, we find that $\frac{\bar{\omega}_1^2}{\bar{m}_1^2}\,<\,1$ for $\bar{M}=10$ and $\bar{Q}=7,17$ while  $\frac{\bar{\omega}_1^2}{\bar{m}_1^2}\,>\,1$ for $\bar{M}=10$ and $\bar{Q}=6$, and it seems that $\frac{\bar{\omega}_1^2}{\bar{m}_1^2}\,<\,1$ for all $|\bar{Q}|\,>\,\frac{\sqrt{7}\bar{M}}{4}$ while
 $\frac{\bar{\omega}_1^2}{\bar{m}_1^2}\,>\,1$ for all $|\bar{Q}|\,<\,\frac{\sqrt{7}\bar{M}}{4}$.
 Both of Figure \ref{fig:1} and Figure \ref{fig:2} are examples of scalar field profiles with $\frac{\omega}{m}\,<\,1$ (as for $\phi$ particles). The existence of the wave profiles of the form of the figures \ref{fig:1}, \ref{fig:2} is consistent with the accumulation of the scalar particles at some distance from the black hole that is suggested by the test particle behaviour predicted in the first part of the preceding section. We observe that $\frac{\omega}{m}\,>\,1$ always correspond to imaginary $\omega$'s, so the corresponding solutions are unstable. It is also observed that for some $M$, $Q$ pairs the absolute value of $\psi_\omega$ peaks at some values of Q, for example, as in Figure \ref{fig:3}.

 We notice that the $\frac{f^\prime}{r}$ term is negligible with respect to $m_\chi^2$ for phenomenologically viable values of the parameters as can be seen below
\begin{equation}
\frac{f^\prime}{\bar{r}}= \frac{2}{\bar{r}^2}\left(\frac{\bar{M}}{\bar{r}}-\frac{\bar{Q}^2}{\bar{r}^2}\right),
\label{c4x}
\end{equation}
 where we have replaced $r$, $M$, $Q$ by their barred forms and using  (\ref{bar1})-(\ref{bar6}). We find that for values of $r$ greater than $r_+$, $\bar{r}$ is at least at the order of $\frac{M}{M_\circ}$, $\bar{M}$ is $\frac{M}{M_\circ}$, $\bar{Q}$ for black hole solutions is at most in the order of $\frac{M}{M_\circ}$. Therefore, $\frac{f^\prime}{r}$ is at most in the order of 1 and for most values of $r$ it is much smaller than 1. It is evident from (\ref{bar3}) that $\bar{m}\,>\,1$ for $mc^2\,>\,10^{-9}eV$. This, in turn, implies that one get good information about $\psi_\omega$ in the 3+1 dimensional case for $l=0$ (i.e. for radial motion) by studying $\psi_\omega$ given in this study provided that either $mc^2\,\gg\,10^{-9}eV$ or $\bar{r}\,\gg\,1$. There may be also situations where $mc^2\,>\,10^{-9}eV$ and $\bar{r}\,>\,2$ and $\psi_\omega$ for $l=0$ is a good approximation to (\ref{c2}). On the other hand, we have found above that the phenomenologically viable values of particle masses in this setup satisfy $mc^2\,<\,10^{-9}eV$. However, the phenomenologically relevant interval $\{$$|\bar{Q}|\,>\,\frac{\sqrt{7}\bar{M}}{4}$, $\bar{m}_1^2$, $\pm\bar{\omega}_1$, $\pm\bar{q}_1$$\}$ obtained above includes the case where $|\bar{Q}|\sim\,\bar{M}$. Note that for small values of $\bar{r}$ outside the horizon $r_+$, we have $\frac{\bar{M}}{\bar{r}}\,\simeq\,1$. The case $|\bar{Q}|\sim\,\bar{M}$ may make $\frac{f^\prime}{r}$ negligible with respect to $m_\chi^2$ even for $mc^2\,<\,10^{-9}eV$ for most of the values of $\bar{r}$ since the $\left(\frac{\bar{M}}{\bar{r}}-\frac{\bar{Q}^2}{\bar{r}^2}\right)$ term ensures $\frac{f^\prime}{\bar{r}}\sim\,0$ for small values of $\bar{r}$ that are in the order of 1 while the $\frac{2}{\bar{r}^2}$ term ensures $\frac{f^\prime}{\bar{r}}\sim\,0$ for large values of $\bar{r}$.

 We have found that the $\phi$ and $\chi$ fields and the black hole must have dark electric charges. In this study we have considered small energy densities of $\phi$ and $\chi$ fields so that they do not change the geometry of the space.  Therefore, it is quite difficult to detect these dark matter candidates. On the other hand, we do not expect a radical change in the form of the geometry even when the energy density of the fields is increased provided we are at a sufficiently large distance from the black hole so that (\ref{c4}) is satisfied and the spherical shape of the wave profile is preserved i.e. $l=0$. In that case the geometry of the compact object will be still described by the Reissner-Nordstr{\o}m metric. In such a situation, the presence of the scalar fields charged with a dark electric charge around a Reissner-Nordstr{\o}m black hole (charged with the same dark electric charge) can be detected by the gravitational effect of these field(s) e.g. through their effect on the rotation curve(s) of their galax(ies) (while such an analysis will have additional, non-trivial points to be addressed). All these points need a separate and detailed analysis. To reach a definite and rigorous conclusion for the effect of non-negligible energy densities of the scalar fields, all these points must be considered in rigorous, separate detailed future studies.

\section{conclusion}

In this study we considered the problem of evolution of a heavier scalar field $\phi$ that is produced from a lighter homogeneously distributed scalar field $\chi$ through a $\chi^*\chi\phi^*\phi$ interaction term in the background of a Reissner-Nordstr{\o}m black hole. To see the situation better, first, we have studied the problem at a wholly classical setting at the level of test particles. We have observed that $\phi$ particles tend to accumulate at some distance from the black hole which provides a suitable condition for condensation. Then, we have considered the problem at the framework of field theory. We have found approximate solitonic-like solutions for the scalar fields where the heavier $\phi$ particles seem to be more localized compared to the lighter $\chi$ partiles. This wave profile seems to suggest suitable conditions for condensation as in the case of the wholly classical treatment. We have also discussed the phenomenological viability of this model. The requirement of phenomenological viability of the model suggests that the black hole and the scalar particles should have a dark U(1) charge rather than the usual electromagnetic charge to sustain the soliton-like configuration studied in this study in a realistic framework. Note that we have argued that the $\phi$ field is produced from $\chi$ field through $\chi^*\chi\phi^*\phi$ interactions while we have neglected interactions between the $\phi$ and $\chi$ as we have obtained the wave profiles of the solutions. Although this approach may be considered as a sufficiently good approximation for small coupling constant $\lambda$, a separate study in future where this interaction is not neglected in the derivation of the wave profile would be useful to understand all aspects of the problem.

The prospect of studying the extensions of this model along the lines mentioned above seems promising. It is a well-known fact that, in view of rotation curves of spiral galaxies and other astronomical data, there should be a localized distribution of dark matter around the centers of these galaxies. Moreover, many of such galaxies contain supermassive black holes at their centers. Therefore, the model discussed in this study has the potential to describe such localized distributions of dark matter after the model is extended to the case of non-negligible energy density for the $\phi$ fields provided that (at least some of) the supermassive black holes may be identified by the type of black holes discussed in this study. It will be interesting to study these points in detail in future.

\begin{acknowledgments}
%We would like to thank Professor Vitor Cardoso for reading the manuscript and for his valuable comments.

This paper is financially supported by {\it The Scientific and Technical Research Council of Turkey (T{\"{U}}BITAK)} under the project 117F296 in the context of the COST action {\it CA 16104 "GWverse"}
\end{acknowledgments}

\appendix

\section{The scalar field equation in 1+1 dimensions}

In this appendix we show that the approximate scalar field equation corresponding to  $\frac{f^\prime}{\bar{r}}\simeq\,0$, namely, the equation (\ref{eq:15}) is the scalar field equation in 1+1 dimensions.

We consider the following action for $\chi$ and $\phi$ particles
\begin{eqnarray}
S
&=&\int\;dt\,dr\,\{-g^{\mu\nu}\left[D_\mu\phi\,\left(D_\nu\phi\right)^*\,+\,
D_\mu\chi\,\left(D_\nu\chi\right)^*\right]\,-\,m_\phi^2\left|\phi\right|^2\,-\,
m_\chi^2\left|\chi\right|^2\,-\,\lambda\,\phi^*\phi\chi^*\chi\}, \nonumber \\
&&\label{eq:9aap}
\end{eqnarray}
where $D_\mu\,=\,\partial_\mu\,+\,iq\,A_\mu$ with $q$ being the electric charge of the scalar field and $A_\mu\,=\,\left(\frac{Q}{r}, 0, 0, 0\right)$ denoting the electric field of the black hole. We let both $\chi$ and $\phi$ have the same charge q. After the change of variables $dr_*\,=\,f^{-1}\,dr$, (\ref{eq:9aap}) becomes
\begin{eqnarray}
S\,=\,
\int&dt\,dr_*&\{\left|\frac{\partial\phi}{\partial\,t}+i\frac{qQ}{r}\right|^2-\left|\frac{\partial\phi}{\partial\,r_*}\right|^2 -\tilde{m}_\phi^2\phi^2 \nonumber \\
&&+\,\left|\frac{\partial\chi}{\partial\,t}+i\frac{qQ}{r}\right|^2-\left|\frac{\partial\chi}{\partial\,r_*}\right|^2
-\tilde{m}_\chi^2\chi^2
-\tilde{\lambda}\phi^*\phi\chi^*\chi\},
\label{eq:9bap}
\end{eqnarray}
where
\begin{equation}
\tilde{m}_i^2\,=\,f\,m_i^2~,\tilde{\lambda}\,=\,f\,\lambda~~~~i=\chi,\phi.  \label{eq:10}
\end{equation}
Transforming (\ref{eq:9aap}) to (\ref{eq:9bap}) corresponds to changing the metric $ds^2$ into $d\tilde{s}^2$ where $ds^2$ and $d\tilde{s}^2$ are related by
\begin{equation}
ds^2\,=\,f\left(-\,dt^2\,+\,f^{-2}dr^2\right)\,=\,f\,d\tilde{s}^2, \label{eq:11ap}
\end{equation}
where
\begin{equation}
d\tilde{s}^2\,=\, -\,dt^2\,+\,dr_*^2 \label{eq:12ap}.
\end{equation}
In other words we have passed to an effective Minkowski space given by (\ref{eq:12ap}) at the expense of making the masses and the coupling constant r-dependent (hence,$r_*$-dependent).

In the following we obtain the approximate profile of the distribution of the scalar particles $\chi$ and $\phi$. In (\ref{eq:9aap}) and (\ref{eq:9bap})we have neglected the effect of electromagnetic interactions between the scalar particles since the coupling constant of electromagnetic interactions is small, and the density of the scalar particles are taken to be small. In a similar way we  take the coupling constant $\lambda$ to be small. Then the approximate field equation corresponding to (\ref{eq:9bap}) for $\phi$ is
\begin{equation}
\tilde{D}_\mu\,\tilde{D}^\mu\,\phi\,-\,\tilde{m}_\phi^2\,\phi\,=\,\frac{\partial^2\phi}{\partial\,t^2}\,-\,\frac{\partial^2\phi}{\partial\,r_*^2}
\,+\,\frac{2iqQ}{r}\frac{\partial\phi}{\partial\,t}\,+\,\left(\tilde{m}_\phi^2\,-\,\frac{q^2Q^2}{r^2}\right)\phi\,=\,0,
\label{eq:13ap}
\end{equation}
where $\tilde{D}_\mu=\tilde{\partial}_\mu+iqA_\mu$$=$$\,\left(\frac{\partial}{\partial\,t}+i\frac{qQ}{r}, \frac{\partial}{\partial\,r_*}\right)$.
The corresponding equation for $\chi$ may be obtained by replacing $\phi$ in (\ref{eq:13ap}) by $\chi$.

The total mechanical energy for the metric (\ref{eq:3}) (i.e. for the local effective Minkowski space) $\tilde{E}$ is equal to the total energy of the particle. In other words
$\tilde{E}_i^2\,=\,\tilde{p}_i^2+\tilde{m}_i^2\,=\,m_i^2\,C_i^2$. Hence the oscillatory solutions may be taken as
\begin{equation}
\phi\,=\,R_\phi(r_*)\,e^{-i\omega_\phi\,t},
\label{eq:14ap}
\end{equation}
where $\omega_\phi$ is identified by $\tilde{E}=\hbar\omega_\phi=\hbar\,m_\phi\,C_\phi$. Thus, (\ref{eq:13ap}) reduces to
\begin{equation}
\frac{d^2R}{dr_*^2}\,+\,\left[\left(\omega-\frac{qQ}{r}\right)^2\,-\,\tilde{m}_\phi^2\right]\,R\,=\,0,
\label{eq:15ap}
\end{equation}
which is the same as (\ref{eq:15}).

\section{Derivation of the equations (\ref{H5})-(\ref{H9})}

Consider (\ref{H2}), namely,
\begin{align}
\frac{g^{\prime\prime}}{g}+2is\frac{g^{\prime}}{g}-\frac{2(qQ\omega-m^2M)}{r}+\frac{(q^2-m^2)Q^2}{r^2}=0,\label{H2ap}
\end{align}
where $\prime$ denotes the derivative with respect to $r_*$.

We let
\begin{align}
\frac{g^{\prime}}{g}=\frac{\alpha_1}{r}+\frac{\beta_1}{r^2}+\frac{\gamma_1}{r^3}, \label{H3apa}
\end{align}
\begin{align}
\frac{g^{\prime\prime}}{g}=\frac{\alpha_2}{r}+\frac{\beta_2}{r^2}+\frac{\gamma_2}{r^3}. \label{H3apb}
\end{align}
Next, we use (\ref{H3apa}), (\ref{H3apb}) and $^\prime=\frac{d}{dr_*}=\frac{dr}{dr_*}\frac{d}{dr}$ and the following identity to relate $\alpha_1$, $\beta_1$, $\gamma_1$ and $\alpha_2$, $\beta_2$, $\gamma_2$
\begin{align}
\frac{d}{dr_*}\left(\frac{g^{\prime}}{g}\right)
= \left(\frac{g^{\prime}}{g}\right)^\prime= \frac{g^{\prime\prime}}{g} - \left(\frac{g^{\prime}}{g}\right)^2. \label{H3apx}
\end{align}
Hence, we obtain
\begin{align}
\frac{g^{\prime\prime}}{g}&=\frac{1}{r^2}\alpha_1(\alpha_1-1)+\frac{1}{r^3}2(M\alpha_1-\beta_1+\alpha_1\beta_1)+\frac{1}{r^4}(4\beta_1 M- Q^2\alpha_1 - 3\gamma_1 +2\alpha_1 \gamma_1 +\beta_1^2) \nonumber
\\&
+\frac{1}{r^5}2(3M\gamma_1-Q^2\beta_1+\beta_1\gamma_1)+\gamma_1(\gamma_1-3Q^2)\frac{1}{r^6}=\frac{\alpha_2}{r}
+\frac{\beta_2}{r^2}+\frac{\gamma_2}{r^3}.  \label{H4ap}
\end{align}
(\ref{H4ap}) implies
\begin{eqnarray}
&&\alpha_2=0\;~,~~\beta_2=\alpha_1(\alpha_1-1)\;~,~~\gamma_2=2(M\alpha_1-\beta_1+\alpha_1\beta_1)\;,\nonumber \\
&&4M\beta_1-Q^2\alpha_1-3\gamma_1+\beta_1^2+2\gamma_1\alpha_1=0 \;~,~~
3M\gamma_1-Q^2\beta_1+\beta_1\gamma_1=0\;~,~\gamma_1-3Q^2=0. \nonumber \\
\label{H4apx}
\end{eqnarray}
(\ref{H4apx}) reults in
\begin{equation}
\alpha_1=\frac{9}{5}-\frac{9}{20}\bigg(\frac{M}{Q}\bigg)^2 \ \ \ \ \ \ \ \beta_1=-\frac{9}{2}M \ \ \ \ \ \ \ \gamma_1=3Q^2 \label{H5ap}
\end{equation}
\begin{align}
\alpha_2=0 \,, \ \ \ \ \ \ \beta_2=\bigg[\frac{9}{5}-\frac{9}{20}\bigg(\frac{M}{Q}\bigg)^2\bigg]\bigg[\frac{4}{5}-\frac{9}{20}\bigg(\frac{M}{Q}\bigg)^2\bigg] \,, \ \ \ \ \ \ \gamma_2=2\bigg[\frac{-18}{5}+\frac{63}{40}\bigg(\frac{M}{Q}\bigg)^2 \bigg]M.  \label{H6ap}
\end{align}
Inserting solutions (\ref{H5ap}) and (\ref{H6ap}) into (\ref{H2}) and using $s^2=\omega^2-m^2$, we get three equations for three unknown quantities $w$, $q$, $m$ for a given $M$ and $Q$:
\begin{align}
(m^2-\omega^2)-\frac{M^2}{9Q^4}\bigg(\frac{18}{5}-\frac{63}{40}\frac{M^2}{Q^2}  \bigg)^2=0,
\end{align}
\begin{align}
-\frac{3M^2}{Q^2}\bigg(\frac{18}{5}-\frac{63}{40}\frac{M^2}{Q^2}\bigg)+\bigg(\frac{9}{5}-\frac{9}{20}\frac{M^2}{Q^2} \bigg) \bigg( \frac{4}{5}-\frac{9}{20}\frac{M^2}{Q^2} \bigg) +(q^2-m^2)Q^2=0,
\end{align}
\begin{align}
\frac{2M}{3Q^2}\bigg(\frac{18}{5}-\frac{63}{40}\frac{M^2}{Q^2}\bigg)\bigg(\frac{9}{5}-\frac{9}{20}\frac{M^2}{Q^2}\bigg)-2(qQ\omega-m^2M)=0.
\end{align}
These three equations solve $m$, $\omega$, and $q$ as given in (\ref{eq:m2})-(\ref{eq:q22}).

\section{Derivation of (\ref{scaled_eq})}

After inserting the Newton's constant $G$, the speed of light $c$, the Planck's constant $\hbar$ into explicitly, the action for a charged free scalar field $\phi$ becomes
\begin{align}
S=\int \pmb{\hbar^2}\bigg[-g^{\mu\nu} (D_\mu \phi) ( D_\mu \phi)^*-\frac{\pmb{m^2c^2}}{\pmb{\hbar^2}}\phi\phi^*\bigg]d^2x.
\end{align}
Note that we are neglecting possible interaction terms other than electromagnetic interactions because we take other possible interactions negligible with respect to the other terms in the Lagrangian.
The corresponding equation is
\begin{equation}
\,\frac{\partial^2\phi}{\pmb{c^2}\partial\,t^2}\,-\,\frac{\partial^2\phi}{\partial\,r_*^2}\,+ \frac{2i\pmb{k}}{\pmb{\hbar}\pmb{c^2}}\frac{\pmb{qQ}}{r}\frac{\partial\phi}{\partial\,t}\,+\,\left(\frac{\pmb{c^2}}{\pmb{\hbar^2}}\tilde{\pmb{m}}^2_\phi
\,-\,\frac{\pmb{k^2}}{\pmb{\hbar^2} \pmb{c^2}}\frac{\pmb{q}^2 \pmb{Q}^2}{r^2}\right)\phi\,=\,0,
\label{eq:13apx}
\end{equation}
where $\pmb{k=\frac{1}{4\pi\epsilon_0}}$.

After inserting (\ref{c1}) into (\ref{eq:13apx}) and using (\ref{c4}), we get
\begin{equation}
\frac{d^2\psi_\omega}{dr_*^2}\,+\,\left[\left(\frac{\pmb{\omega}}{\pmb{c}}-\frac{\pmb{kqQ}}{\pmb{\hbar c} r}\right)^2\,-\,\frac{\pmb{c^2}}{\pmb{\hbar^2}}\pmb{m}^2_\phi\left(1-\frac{2\pmb{GM}}{\pmb{c^2}r}+\frac{\pmb{kGQ^2}}{r^2\pmb{c^4}}\right)\right]\,\psi_\omega\,=\,0.
\label{eq:15'ap}
\end{equation}

To express $r_*$ in terms of the Schwarzschild radius of the sun we multiply both sides of (\ref{eq:15'ap}) by $\left(\frac{GM_\circ}{c^2}\right)^2$.
Then (\ref{eq:15'ap}) becomes
\begin{align}
\frac{d^2\psi_\omega}{d\overline{r}_*^2}+\bigg[\bigg(\frac{\omega GM_\circ}{c^3}- \frac{kqQ}{\hbar c \overline{r}} \bigg)^2-m_\phi^2\bigg(\frac{GM_\circ}{\hbar c} \bigg)^2\left(1-\frac{2M}{M_\circ \overline{r} } + \frac{Q^2 k}{M_\circ^2 G \overline{r}^2}\right)\bigg]\psi_\omega=0,
\end{align}
where
\begin{align}
\overline{r}_*=\frac{c^2}{GM_\circ}r_*,~~ \overline{r}=\frac{c^2}{GM_\circ}r.
\end{align}
We may define
\begin{align}
&\overline{\omega}=\omega\frac{GM_\circ}{c^3}=\bigg(\frac{\omega}{s^{-1}}\bigg)5\times 10^{-6},
\\&
\overline{m}=m\frac{GM_\circ}{\hbar c} =\bigg( \frac{m}{kg}\bigg)\,4.5\times 10^{45},
\\&
\overline{M}=\frac{M}{M_\circ}=\bigg(\frac{M}{kg}\bigg)2\times 10^{-30},
\\&
\overline{Q}=\frac{Q\sqrt{k}}{M_\circ \sqrt{G}}=\bigg(\frac{Q}{C}\bigg)5,7\times 10^{-21},
\end{align}
\begin{align}
\frac{kqQ}{\hbar c}=\overline{q}\overline{Q}=\frac{q\overline{Q}M_\circ \sqrt{Gk}}{\hbar c } \ \ \ \ \Rightarrow \ \ \ \ \overline{q}=q\frac{M_\circ \sqrt{Gk}}{\hbar c}=\bigg(\frac{q}{C}\bigg) 4,8 \times 10^{55},
\end{align}
where we have used the numerical values of $M_\circ$, $G$, $c$, $\hbar$ in SI unit system.

Then, (\ref{eq:15'ap}) becomes
\begin{align}
\frac{d^2\psi_\omega}{d\overline{r}_*^2}+\bigg[\big(\overline{\omega}-\frac{\overline{q}\overline{Q}}{\overline{r}}\big)^2-\overline{m}_\phi^2\left(1-\frac{2\overline{M}}{\overline{r}} + \frac{\overline{Q}^2}{\overline{r}^2}\right) \bigg]\psi_\omega=0.
\label{scaled_eqap}
\end{align}

\end{document}